\documentclass[12pt]{article}

\begin{document}

\title{\addvspace{-15mm} {\normalsize \hfill CFNUL/99-08 }\\
{\normalsize \hfill hep-ph/9908506}\\
\addvspace{20mm}
Classical Nambu-Goldstone fields\footnote{Talk given at the X International 
School "Particles and Cosmology", 19-25 April, 1999, Baksan Valley,
Kabardino-Balkaria, Russia}}
\author{Lu\'{\i }s Bento \\
{\normalsize {\em Centro de F\'{\i }sica Nuclear da Universidade de Lisboa,}}%
\\
{\normalsize {\em Av. Prof. Gama Pinto 2, 1649-003 Lisboa, Portugal}}}
\date{}
\maketitle

\begin{abstract}
It is shown that a true Nambu-Goldstone (NG) boson develops a coherent
long-range field whenever the charge associated with it that is carried by
the other particles is not conserved in a macroscopic scale. The source of a
NG field is the time rate of quantum number violation. If the lepton numbers
are spontaneously broken at a scale below 1 TeV, the neutrino oscillation
processes generate long-range majoron fields that are strong enough in
Supernovae to modify the neutrino flavor dynamics. Two examples are given:
NG fields may improve the adiabaticity of $\nu _{e}\leftrightarrow \nu _{X}$
transitions or cause {\em resonant} anti-neutrino oscillations otherwise
impossible with solely weak interactions.
\end{abstract}

\newpage
\section{Basic Features}

The Goldstone\ theorem asserts that in a theory with spontaneous breaking of
a global symmetry group some massless scalar bosons should exist -- the
so-called Nambu-Goldstone bosons -- one per broken global symmetry. Good
candidates for spontaneously broken quantum numbers are the partial lepton
numbers $L_{e},L_{\mu },L_{\tau }$ and total lepton number $L$. Indeed, they
are conserved by the Standard Model (SM) interactions, but several
experiments going now from solar to atmospheric neutrinos and laboratory
oscillation experiments \cite{cald98,naka99} indicate that the lepton
flavors are not conserved.

The NG bosons not only have zero mass but also no scalar potential terms
such as $\phi ^{4}$. They interact with the fermion particles however, do
not mediate long range forces. The reason lies in the very global symmetry
which, broken by the vacuum still operates at the Lagrangian level, realized
as a translation of the NG field, $\phi \rightarrow \phi +\alpha \,,$ $%
\alpha ={\rm constant}$. For that reason they only have derivative couplings 
\cite{gelm83} {\em i.e.}, the Lagrangian only depends on the derivatives $%
\partial _{\mu }\phi $. The $\phi $ equation of motion identifies with the
conservation law associated to the $\phi \rightarrow \phi +\alpha \,$
symmetry,
\begin{equation}
\partial _{\mu }\partial ^{\mu }\,\phi =-\partial _{\mu }J_{\Lambda }^{\mu
}/V_{\Lambda }\,,  \label{ddfi}
\end{equation}
where $V_{\Lambda }$ is essentially the scale of symmetry breaking and the
current $J_{\Lambda }^{\mu }$ is determined in leading order by the quantum
numbers of the other particles. For instance, if $\Lambda $ is one of the
lepton numbers $L_{e}$ or $L_{\mu }$, the current and interaction Lagrangian
are given by
\begin{eqnarray}
{\cal L}_{{\rm int}} &=&J_{\Lambda }^{\mu }\,\partial _{\mu }\phi
/V_{\Lambda }+\mbox{\rm quadratic terms in }\partial _{\mu }\phi \,\;,
\label{lngint} \\
J_{\Lambda }^{\mu } &=&\Lambda _{e}({\bar{e}\,}\gamma ^{\mu }\,e+{\bar{\nu}%
_{e}}\gamma ^{\mu }\nu _{e})+\Lambda _{\mu }({\bar{\mu}\,}\gamma ^{\mu
}\,\mu +{\bar{\nu}_{\mu }}\gamma ^{\mu }\nu _{\mu })+\cdots \;\,,
\label{jll}
\end{eqnarray}
where the dots stand for model dependent scalar boson contributions and
higher order effective couplings.

All this can be easily understood \cite{gelm83,bent97}. Let $\Lambda _{a}$
and $\Lambda _{i}$ be the quantum numbers of the fermion fields $\chi ^{a}$
and scalars $\sigma _{i}$ under an abelian global symmetry${\rm {\
U(1)_{\Lambda }}}$ which is spontaneously broken by vacuum expectation
values $\left\langle \sigma _{i}\right\rangle $. A suitable change of
variables namely,
\begin{eqnarray}
\chi ^{a} &=&e^{-i\,\Lambda _{a}\phi /V_{\Lambda }}\psi ^{a}\ ,\smallskip 
\label{tpsi} \\
\sigma _{i} &=&e^{-i\,\Lambda _{i}\phi /V_{\Lambda }}\,\left( \left\langle
\sigma _{i}\right\rangle +\rho _{i}\right) \ ,  \label{tsigma}
\end{eqnarray}
makes the Lagrangian to be expressed in terms of {\em physical weak}
eigenstates, fermions $\psi ^{a}=e,\,\nu _{e},\,...$, massive bosons $\rho
_{i}$ and gauge bosons (assumed to be singlets of the global symmetry) all
of them invariant under U(1)$_{\Lambda }$. The only non-invariant field is
the NG boson $\phi $. Its couplings are derived from the kinetic Lagrangians
of $\chi ^{a}$ and $\sigma _{i}$ by application of the transformations (\ref
{tpsi}-\ref{tsigma}).

The second member of the NG equation of motion, Eq.\ (\ref{ddfi}), is
clearly not an ordinary scalar density. If one calculates the divergency of
a vector or axial-vector fermion current using the Dirac equation, one ends
up with pseudo-scalars (both diagonal and non-diagonal in flavor) and {\em %
off-diagonal} scalar densities: 
\begin{eqnarray}
&&\partial _{\mu }\,\overline{f}_{i}\gamma ^{\mu }\gamma
_{5}f_{j}=i(m_{i}+m_{j})\overline{f}_{i}\gamma _{5}f_{j}\ ,  \label{f5f} \\
&&\partial _{\mu }\,\overline{f}_{i}\gamma ^{\mu }f_{j}=i(m_{i}-m_{j})%
\overline{f}_{i}f_{j}\;.  \label{ff}
\end{eqnarray}
However couplings to {\em flavor-diagonal} scalar densities are not possible 
\cite{gelm83,chik81}. The pseudo-scalars that are diagonal in flavor vanish
for free particle states and the flavor violating densities depend on the
relative phase of distinct flavors. The natural conclusion has been that
these relative phases cancel each other when summed over a large number of
particles and therefore long range '$1/r$' NG fields are not possible, only 
{\em spin-dependent} '$1/r^{3}$' potentials can exist \cite{gelm83,chik81}.
It will be clear in a moment that this is not true \cite
{bent97,bent98,trento98,zurab99}.

\section{Long range Nambu-Goldstone fields}

First notice that the source of a NG field does not vanish if the matter
current $J_{\Lambda }^{\mu }$ is not conserved. The volume integral 
\begin{equation}
\int \!{d^{3}x\,\ }\partial _{\mu }J_{\Lambda }^{\mu }=\frac{d\Lambda }{dt}^{%
\hspace{-2.83pt}{\rm created}}  \label{dLdt}
\end{equation}
represents the rate of {\em creation} per unity of time of the $\Lambda $%
-number carried by all particles except $\phi $. This rate, divided by $%
-V_{\Lambda }$, constitutes the source of the NG boson and 
\begin{equation}
\phi (t,\vec{r})=\frac{-1}{V_{\Lambda }}\int \!d^{3}x\,{\frac{\partial _{\mu
}J_{\Lambda }^{\mu }(t-\left| {\vec{r}-\vec{x}}\right| ,\vec{x})}{4\pi
\left| {\vec{r}-\vec{x}}\right| }}\;  \label{field}
\end{equation}
is a solution of the equation of motion (\ref{ddfi}). It is a long-range
field but suppressed by the ratio of the rate of $\Lambda $-violating
processes per unity of time over the scale ${V_{\Lambda }}$. To produce a
significant macroscopic field one needs thus a large rate of reactions. It
calls for astrophysical sources. There is evidence that solar neutrinos
change flavor as they come out of the Sun. If $\nu _{e}$ oscillates into ${%
\nu _{\mu }}$, for instance, then ${L_{e}}$ and ${L_{\mu }}$ are not
conserved in a large scale in stars and presumably in Supernovae as well. If
any of these quantum numbers is associated with a spontaneously broken
global symmetry then, the respective NG bosons (so-called majorons \cite
{chik81}) acquire classical long-range field configurations.

For definiteness suppose that $\Lambda =L_{e}-L_{\mu }$ is spontaneously
broken and that a fraction of $\nu _{e}$s emitted from a star are resonantly
converted into $\nu _{\mu }$ through the Mikheyev-Smirnov-Wolfenstein (MSW)
mechanism \cite{wolf78} (also $\nu _{\mu }\rightarrow \nu _{e}$ in a
Supernova) in a certain shell inside the star. For simplicity assume
spherical symmetry and stationary neutrino fluxes in which case the $%
L_{e}-L_{\mu }$ number carried by electrons and neutrinos decreases at a
constant rate in time. The rate $\dot{L}_{e}-\dot{L}_{\mu }$ generates a
long-range NG field of Coulombian form, charge divided by $4\pi r$, above
the resonance shell, 
\begin{equation}
\phi =\frac{1}{4\pi r}\frac{2}{V_{\Lambda }}\left[ \dot{N}({\nu
_{e}\rightarrow \nu _{\mu })-}\dot{N}({\nu _{\mu }\rightarrow \nu _{e})}%
\right] \;,  \label{phi}
\end{equation}
whereas $\phi $ is constant below the resonance shell. This long-range '$1/r$%
' field configurations are in conflict with the conclusion traditionally
taken by looking at the fermion bilinears (\ref{f5f}-\ref{ff}) that appear
as a source of a NG boson namely, that only '$1/r^{3}$' spin-dependent NG
potentials could exist. This particular issue is the subject of a recent
work \cite{zurab99} and is examined below.

In the interest of clearness let us concentrate on a specific system, the $%
\nu _{e}-\nu _{\mu }$ pair of flavors. Their interactions are assumed to be
weak interactions with a background medium, couplings with a NG boson
associated to $\Lambda =L_{e}-L_{\mu }$ and a $2\times 2$ {\em real}
Majorana mass matrix $m$, all contained in: 
\begin{eqnarray}
{\cal L}_{{\rm int}} &=&-\sqrt{2}G_{{\rm F}}n_{e}(\bar{\nu}_{e}\gamma
^{0}\nu _{e})-\frac{g\langle Z^{0}\rangle }{2\cos \theta _{{\rm W}}}(\bar{\nu%
}_{e}\gamma ^{0}\nu _{e}+\bar{\nu}_{\mu }\gamma ^{0}\nu _{\mu })  \nonumber
\\
&+&\frac{1}{V_{\Lambda }}\partial _{\mu }\phi (\bar{\nu}_{e}\gamma ^{\mu
}\nu _{e}-\bar{\nu}_{\mu }\gamma ^{\mu }\nu _{\mu })-(\nu _{L}^{T}Cm\nu _{L}+%
\mbox{\rm H. C.})\;.  \label{lint}
\end{eqnarray}
$n_{e}$ is the electron density. Notice that only $m_{ee}$ and $m_{\mu \mu }$
break $L_{e}-L_{\mu }$. This Lagrangian determines the equations of motion
of the (left-handed) neutrino operators and also of the single particle wave
functions, related to them by a relation of the type $\nu _{L}+\nu _{L}^{C}=$
$\int \!a\,\psi +a^{\dagger }\,\psi ^{C}$. The wave functions obey the
equations 
\begin{equation}
i\,\partial \hspace{-0.22cm}/\psi _{a}=m_{ab}\,\psi _{b}-V_{a}^{\mu
}\,\gamma _{\mu }\gamma _{5}\psi _{a}\;,  \label{dpsi}
\end{equation}
where $a$ stands for $\nu _{e}$ and $\nu _{\mu }$ and $V_{a}^{\mu }$
represent the {\em flavor conserving} weak and majoron vector potentials ($%
V_{a}^{\mu }=-\Lambda _{a}\partial ^{\mu }\phi /V_{\Lambda }$ in the latter
case). When $\partial _{\mu }J_{\Lambda }^{\mu }$ is evaluated over a system
of neutrino particles the equation of motion (\ref{ddfi}) becomes \cite
{zurab99}: 
\begin{equation}
\partial _{\mu }\partial ^{\mu }\phi =\frac{2}{V_{\Lambda }}\sum_{\nu
}i\left( m_{ee}\,\bar{\psi}_{\nu _{e}}\gamma _{5}\psi _{\nu _{e}}-m_{\mu \mu
}\,\bar{\psi}_{\nu _{\mu }}\gamma _{5}\psi _{\nu _{\mu }}\right) \;.
\label{ddfiwf}
\end{equation}
As expected, the parameters $m_{ee}$ and $m_{\mu \mu }$ that break $%
L_{e}-L_{\mu }$ in the Lagrangian (\ref{lint}) appear explicitly in the
source terms. These are precisely the kind of pseudo-scalar densities
identified in Eq.~(\ref{f5f}). They would vanish identically if the
neutrinos were free particles. But they are not. When studying neutrino
oscillations one usually separates the spin from the flavor degrees of
freedom by expressing the wave function as a product of a left-handed {\em %
free massless} spinor $\psi _{0}^{\alpha }$ ($\gamma _{5}\psi _{0}=-\psi _{0}
$, $i\,\partial \hspace{-0.22cm}/\psi _{0}=0$) and a flavor-valued wave
function $\varphi _{a}$. It obeys a well known evolution equation \cite
{kuo89} as a function of the distance $s$ travelled by each particle, 
\begin{equation}
i\frac{d}{ds}\left( 
\begin{array}{c}
{\varphi _{e}} \\ 
\\ 
{\varphi _{\mu }}
\end{array}
\right) =\frac{1}{2E}\left( 
\begin{array}{cc}
m_{ee}^{2}+2EV_{e} & \quad \frac{1}{2}\Delta m^{2}\sin 2\theta  \\ 
&  \\ 
\frac{1}{2}\Delta m^{2}\sin 2\theta  & \quad m_{\mu \mu }^{2}+2EV_{\mu }
\end{array}
\right) \left( 
\begin{array}{c}
{\varphi _{e}} \\ 
\\ 
{\varphi _{\mu }}
\end{array}
\right) \;,  \label{dfids}
\end{equation}
see Eqs.~(\ref{va}-\ref{VW}). However, pure chiral spinors give vanishing
scalar densities in Eq.~(\ref{ddfiwf}) so one has to go beyond the
approximation $\psi _{a}=\psi _{0}\varphi _{a}$.

For a spherical symmetric NG field as in Eq.~(\ref{phi}) the velocity of a
radially moving neutrino is parallel to the vector potentials ${\bf V}%
_{a}\propto {\bf \nabla }\phi $ and in that case an approximate solution of
Eq.~(\ref{dpsi}) is \cite{zurab99} 
\begin{equation}
\psi _{a}^{\alpha }(x)\cong \psi _{0}^{\alpha }\,\varphi _{a}+\gamma
_{\alpha \beta }^{0}\frac{m_{ab}}{2E}\,\psi _{0}^{\beta }\,\varphi _{b}\;.
\label{psiaa}
\end{equation}
Applying to the majoron equation of motion (\ref{ddfiwf}) it becomes 
\begin{eqnarray}
\partial _{\mu }\partial ^{\mu }\phi  &=&\frac{1}{V_{\Lambda }}\sum_{\nu }%
\frac{m_{e\mu }}{E}(m_{ee}+m_{\mu \mu })\,i\,(\varphi _{e}^{\dagger }\varphi
_{\mu }-\varphi _{\mu }^{\dagger }\varphi _{e})\,\psi _{0}^{\dagger }\psi
_{0}  \nonumber \\
&=&\frac{1}{V_{\Lambda }}\sum_{\nu }\frac{\Delta m^{2}}{2E}\sin 2\theta
\,i\,(\varphi _{1}^{\dagger }\varphi _{2}-\varphi _{2}^{\dagger }\varphi
_{1})\,\psi _{0}^{\dagger }\psi _{0}\;.  \label{ddfif1}
\end{eqnarray}
where $\varphi _{1,2}$ refer to the mass eigenstate basis. It is now clear
that the generation of a NG field requires the interference between
different flavor components of the $\nu $ wave functions. But that simply
means the existence of neutrino oscillations. A crucial fact is that the
source term does not depend on arbitrary initial phases of the $\nu $ wave
functions. Hence, there is no automatic phase cancellation when the sum is
taken over a large number of particles. This of course makes possible to
have sources of macroscopic dimensions.

Making use of Eq.~(\ref{dfids}), the last result can be written in terms of
the probabilities of observing the $\nu _{e}$ and $\nu _{\mu }$ flavors
respectively, $P_{\nu _{e}}=\varphi _{e}^{\dagger }\varphi _{e}\!/\!\varphi
^{\dagger }\varphi $ and $P_{\nu _{\mu }}=1-P_{\nu _{e}}$: 
\begin{equation}
\partial _{\mu }\partial ^{\mu }\phi _{\Lambda }=\frac{-1}{V_{\Lambda }}%
\sum_{\nu }\psi ^{\dagger }\psi \,\frac{d(P_{\nu _{e}}-P_{\nu _{\mu }})}{ds}=%
\frac{-1}{V_{\Lambda }}\frac{d(L_{e}-L_{\mu })}{dt\,d^{3}x}\;.
\label{ddfipe}
\end{equation}
Again, the second member does not depend on the initial phases of the
individual particles. It also tells under which conditions a large number of
particles makes a charge with definite sign: it is when the $\Lambda
=L_{e}-L_{\mu }$ number carried by neutrinos suffers a net increase or
decrease in $\Lambda $-violating processes. We come back to the point of
view at the beginning of this section: the source term of a NG field is
proportional to the rate of {\em creation} (not just local variation) per
unity of time and volume of the quantum number associated to it that is
carried by the matter particles \cite{bent97,bent98,trento98}. In a star or
reactor where only $\nu _{e}$ are produced out of electrons the oscillations 
$\nu _{e}\rightarrow \nu _{\mu }$ necessarily lead to a total decrease of
the lepton number $L_{e}-L_{\mu }$: the electrons captured in nuclear
reactions are, after all, totally or partially transformed into $\nu _{\mu }$
(the probability of observing the $\nu _{\mu }$ flavor is not zero).

\section{Neutrino oscillations: role of NG fields}

Understood as they are the conditions under which a long-range NG field is
produced the next question is what are the implications? A NG field only
interacts through its gradient which implies a suppression factor of $%
V_{\Lambda }^{-1}$ over the time-length scale of field variation. This is
typically an extremely small number. However $\nu $ oscillations are
sensitive to very tiny external potentials that are not independent of the
flavor. This condition is satisfied by the majoron couplings as a rule, see
Eqs.~(\ref{lngint}-\ref{jll}), for example. They contribute with the vector
potentials $V_{a}^{\mu }=-\Lambda _{a}\partial ^{\mu }\phi /V_{\Lambda }$ to
the fermion equations of motion (\ref{dpsi}) and potentials
\begin{equation}
V_{a}=-\Lambda _{a}(\dot{\phi}+{\bf v}\!\cdot \!{\bf \nabla }\phi
)/V_{\Lambda }\;,  \label{va}
\end{equation}
where ${\bf v}$ is the neutrino velocity, to the flavor oscillation equation
(\ref{dfids}).

In the case analyzed in the previous section, the important quantity for $%
\nu _{e}\leftrightarrow \nu _{\mu }$ oscillations is the difference 
\begin{equation}
V_{e}-V_{\mu }=\sqrt{2}G_{{\rm F}}n_{e}+\frac{4}{V_{\Lambda }^{2}}\frac{1}{%
4\pi r^{2}}\left[ \dot{N}(\nu _{e}\rightarrow \nu _{\mu })-\dot{N}({\nu
_{\mu }\rightarrow \nu _{e})}\right] \;.  \label{vevmu}
\end{equation}
It indicates what is the typical order of magnitude of the NG potentials ($%
V_{{\rm NG}}$), neutrino flux over $V_{\Lambda }^{2}$. In the case of the
Sun, $j_{\nu }\!/\!V_{\Lambda }^{2}\sim 10^{-2}/R_{\odot }$ at the Sun
radius for a scale $V_{\Lambda }$ as low as $1\,{\rm KeV}$. It looks too
small to perturb solar neutrino oscillations. In a Supernova however the
fluxes are many orders of magnitude higher and 
\begin{equation}
V_{{\rm NG}}\sim \frac{j_{\nu }}{V_{\Lambda }^{2}}=1.48\ \frac{V_{\Lambda
}^{-2}}{G_{{\rm F}}}\frac{L_{\nu }}{10^{52}\,{\rm ergs/s}}\,\frac{10\,{\rm %
MeV}}{\left\langle E_{\nu }\right\rangle }\left( \frac{r}{10^{10}\,{\rm cm}%
^{{}}}\right) ^{-2}\times {}10^{-12}\,\,{\rm eV}\;,  \label{VNG}
\end{equation}
where $L_{\nu }$ is the $\nu $ energy luminosity, $\left\langle E_{\nu
}\right\rangle $ the average energy and $L_{\nu }\!/\!\left\langle E_{\nu
}\right\rangle $ the neutrino emission rate. In turn, the electroweak
potential \cite{wolf78} is given in a Supernova envelope star by 
\begin{equation}
V_{W}=\sqrt{2}{\rm \,}G_{{\rm F}}\,n_{e}=0.76\ Y_{e}{\frac{{\tilde{M}}}{%
10^{31}{\rm g}}\,}\left( \frac{r}{10^{10}\,{\rm cm}^{{}}}\right) ^{-3}\times
10^{-12}\,\,{\rm eV}\;,  \label{VW}
\end{equation}
where $\tilde{M}=\rho \,r^{3}$ is a constant between 1 and 15 $\times 10^{31}%
{\rm g}$, depending on the star \cite{wils86}.

Clearly, the NG potentials compete with the electroweak potentials for
scales $V_{\Lambda }$ as high as $G_{{\rm F}}^{-1/2}$ and due to their
long-range nature even overcome at large enough radius. One can conceive
that a NG field may affect the propagation of other neutrinos or flavors
outside the region it was created or even outside the star. On the other
hand, it may participate in resonant oscillations for values of $\Delta m^{2}
$ that are interesting for solar or atmospheric neutrino solutions. It was
shown in two previous papers that NG fields may improve the adiabaticity of $%
\nu _{e}\leftrightarrow \nu _{X}$ transitions \cite{bent98} or cause {\em %
resonant} anti-neutrino oscillations otherwise impossible with solely weak
interactions \cite{bent97}.

\subsection{Improvement of adiabaticity: majoron back reaction}

Suppose that $\nu _{e}$ oscillates in the Sun into an active neutrino in
accordance with the non-adiabatic, small mixing angle solution \cite
{haxt86,hata97}. Then, the same $\nu _{e}\rightarrow \nu _{\mu }$
transitions must occur in a Supernova envelope, non-adiabatic as well, but
accompanied by $\nu _{\mu }\rightarrow \nu _{e}$ since $\nu _{\mu }$ are
also produced in a Supernova (for convenience $\nu _{\mu }$ designates the
relevant active neutrino). First, as results from the resonance condition, 
\begin{equation}
2E(V_{e}-V_{\mu })=\Delta m^{2}\cos 2\theta \;,  \label{evevmu}
\end{equation}
the neutrinos with higher energy are converted at a larger radius. Second,
the transitions are less adiabatic and less probable for the higher energy
neutrinos \cite{mina88} which has important consequences: since the $\nu
_{\mu ,\tau }$ flavors are emitted with an hotter spectrum than $\nu _{e}$,
the final observable $\nu _{e}$ spectrum is softer than if the transitions
were completely adiabatic.

The situation changes if there is a majoron field associated to a partial
lepton number, let it be $L_{e}$ \cite{bent98} or $L_{e}-L_{\mu }$ as
considered in the previous sections. The potential $V_{e}-V_{\mu }$ departs
from the charged current potential as shows Eq.~(\ref{vevmu}). The rates $%
\dot{N}(\nu _{e}\rightarrow \nu _{\mu })$ and $\dot{N}(\nu _{\mu
}\rightarrow \nu _{e})$ are functions of the radius, in other words, the $%
\nu $ conversions inside a certain sphere create a NG field
gradient outside that sphere. 
Since a Supernova emits more $\nu _{e}$ than $\nu _{\mu
}$ the majoron potential is positive which has the effect of slowing down
the fall of $V_{e}-V_{\mu }$ with the radius and, by virtue of Eq.~(\ref
{evevmu}), of increasing the radius at which the higher energy neutrinos are
converted. For a fixed $\nu $ energy, the non-adiabaticity increases with
the slope $|d\ln (V_{e}-V_{\mu })\!/\!dr|$ \cite{kuo89,haxt86} thus, the net
effect of the majoron field is to improve the adiabaticity and conversion
efficiency of the hottest neutrinos. The implication, observable in
detectors like Super-Kamiokande and SNO \cite{burr90}, is a $\nu _{e}$
spectrum harder than expected within the framework of electroweak
interactions and non-adiabatic solar neutrino solution.

\subsection{Resonant conversion of anti-neutrinos}

If not only one but all three partial lepton numbers are spontaneously
broken one expects to exist mixing between the three NG bosons $\xi {_{e}}$, 
$\xi {_{\mu }}$, $\xi {_{\tau }}$ \cite{bent97}. By going to the unitary
basis as in Eqs.\ (\ref{tpsi}-\ref{tsigma}), 
\begin{eqnarray}
\chi ^{a} &=&\exp \{-i(\xi {_{e}L}_{e}^{a}+\xi {_{\mu }L}_{\mu }^{a}+\xi {%
_{\tau }L}_{\tau }^{a})\}\,\psi ^{a}\ ,\medskip   \label{tpsi2} \\
\sigma _{i} &=&\exp \{-i(\xi {_{e}L}_{e}^{i}+\xi {_{\mu }L}_{\mu }^{i}+\xi {%
_{\tau }L}_{\tau }^{i})\}\,\left( \left\langle \sigma _{i}\right\rangle
+\rho _{i}\right) \ ,  \label{tsigma2}
\end{eqnarray}
the $\chi ^{a}$ fermion and $\sigma _{i}$ boson kinetic terms deliver the
following interaction Lagrangian 
\begin{equation}
{\cal L}_{{\rm int}}={\frac{1}{2}}V_{\ell m}^{2}\,\partial _{\mu }\xi _{\ell
}\,\partial ^{\mu }\xi _{m}+\partial _{\mu }\xi _{\ell }(\bar{\ell}{\,}%
\gamma ^{\mu }\,\ell +{\bar{\nu}_{\ell }}\gamma ^{\mu }\nu _{\ell
})\;,\qquad \ell ,m=e,\mu ,\tau \;.  \label{lxi}
\end{equation}
The kinetic matrix $V_{\ell m}^{2}$ and its inverse $G_{\ell m}$ (same
dimension as $G_{{\rm F}}$) are not diagonal in general \cite{bent97}. As a
result, long-range majoron fields couple different flavors to each other.
Instead of an equation like (\ref{vevmu}), a neutrino with flavor $\nu
_{\ell }$ feels a NG potential 
\begin{equation}
V_{{\ell }}=-(G_{\ell e}\,\dot{L}_{e}+G_{\ell \mu }\,\dot{L}_{\mu }+G_{\ell
\tau }\,\dot{L}_{\tau })/4\pi r^{2}\;.  \label{vl}
\end{equation}
It means that the violation of one flavor in one place creates majoron
fields that act at a distance on any of the other flavors. This may have
interesting consequences.

Suppose that resonant ${\nu _{e}}\leftrightarrow \nu _{H}${\ oscillations
take place in a Supernova environment (}$H$ stands for Heavy, it may be the
heaviest of the $\nu $ mass eigenstates involved in atmospheric $\nu $
oscillations.) In addition $\nu _{e}$ also mixes with $\nu _{L}$\ ($L$ means
Light, possibly the neutrino involved in solar $\nu $ oscillations) and, {in
absence of other than weak interactions, }$\nu _{e}$ oscillates resonantly
into $\nu _{L}$, but not $\bar{\nu}_{e} \to \bar{\nu}_{L}$, in a region of 
lower density than where ${\nu _{e}}\leftrightarrow \nu _{H}$ take place. 
If majoron fields come into play, the $\nu _{e}$ and $\nu _{L}$
propagation is modified by the potentials in Eq.~(\ref{vl}) that are
 created by the 
${\nu _{e}}\leftrightarrow \nu _{H}${\ oscillations, with }$\dot{L}_{e}=-\,%
\dot{L}_{\mu }<0$ because a Supernova emits more $\nu _{e}$s than $\nu _H$s.
If the following hierarchy, $|G_{e\ell }|$ $<<|G_{LH}|,G_{HH}$, exists between
these constants, the relevant potential for ${\bar{\nu}_{e}}\leftrightarrow {%
\bar{\nu}_{L}}$ oscillations is approximately 
\begin{equation}
V_{\bar{\nu}_{e}}-V_{\bar{\nu}_{L}}=-\sqrt{2}{\rm \,}G_{{\rm F}}\,n_{e}\
-G_{LH}\,\dot{L}_{H}/4\pi r^{2}=-(V_{\nu _{e}}-V_{\nu _{L}})\;.  \label{vevl}
\end{equation}
In absence of NG fields this potential is negative and, by hypothesis, the
resonance condition (\ref{evevmu}) cannot be satisfied for ${\bar{\nu}_{e}}%
\leftrightarrow {\bar{\nu}_{L}}$ contrary to ${\nu _{e}}\leftrightarrow \nu
_{L}$. However, if the majoron fields exist and $G_{LH}$ is negative, the NG
potential is positive and necessarily overcomes at large enough radius
because $n_{e}$ falls faster than $1/r^{2}$. Therefore, if $\Delta m_{eL}^{2}
$\ is sufficiently low and the $\nu $ fluxes high enough resonant
anti-neutrino oscillations become possible. Such effect can in principle be
observed because the predicted energy spectrum of ${\bar{\nu}_{\mu }}$ and ${%
\bar{\nu}_{\tau }}$ is harder than the ${\bar{\nu}_{e}}$ spectrum.

The SN1987A data \cite{hira87} seem to disfavor ${\bar{\nu}_{e}}%
\leftrightarrow {\bar{\nu}_{\mu ,\tau }}$ oscillations \cite{jege96} (see
however \cite{kern95}) but do not exclude that they occur with a probability
below a certain limit. As stressed in \cite{bent97}, they may occur in the first
instants of $\nu $ emission, when the $\nu $ luminosity and hypothetical
majoron fields are stronger, to disappear later below some $\nu $ flux
threshold. This kind of time dependence, if observed, in correlation with
the magnitude of $\nu $ fluxes, would be a clear signature of majoron fields
providing in addition a good measurement of the lepton symmetry breaking.

\section{Conclusions}

In spite of the fact that true Nambu-Goldstone\ bosons only have derivative
couplings it is still possible to obtain macroscopic long-range NG fields.
The source of a NG field is the time rate of decreasing of the quantum
number associated with it that is carried by the matter particles. If the
lepton numbers are spontaneously broken, neutrino oscillations give rise to
long-range majoron fields as a result of the constructive interference
between the wave functions of different flavors. If the scale of lepton
symmetry breaking is at or below the Fermi scale
the associated majoron fields are significant enough
to change the $\nu $ oscillations in Supernovae (SN). 
That corresponds to a sensitivity to effective neutrino-majoron coupling 
constants in scattering processes as low as $m_{\nu }G_{{\rm F}}^{1/2}$!
One of the imprints of
majoron fields is the {\em surprise}: because in solar, atmospheric and
terrestrial neutrino experiments the fluxes are probably too low (or too
high the scale of lepton symmetry breaking) to produce significant majoron
fields, their effects in SN neutrinos cannot be anticipated from those kind
of experiments. The other point is the correlation with the magnitude of $%
\nu $ fluxes. It tells what is the scale of lepton symmetry breaking, and in
a more refined way if, as a result of the few seconds decay of SN $\nu $
emission, the majoron fields and observed SN $\nu $ oscillation patterns
exhibit a time dependence in correlation with the $\nu $ luminosity.\bigskip 

This work was supported in part by the grant PESO/P/PRO/1250/98.


\end{document}